\documentclass[12pt,english,aps,pra,showpacs,onecolumn,tightenlines,groupedaddress,floatfix]{revtex4}

\usepackage{amsmath}
\usepackage{amssymb}
\usepackage{amsfonts}
\usepackage{bbm}
\usepackage{epsfig}
\usepackage{grffile}
\usepackage{babel}

\usepackage{color}
\definecolor{grey}{rgb}{.5,.5,.5}
\definecolor{dblue}{rgb}{0,0,.5}

\usepackage[colorlinks=true,citecolor=dblue,linkcolor=dblue]{hyperref}
\usepackage[all]{hypcap}

\usepackage{setspace}

\newcommand{\id}{\mathbbm{1}}
\newcommand{\Tr}{\operatorname{Tr}}
\newcommand{\bra}{\langle}
\newcommand{\ket}{\rangle}
\newcommand{\mc}[1]{\mathcal{#1}}

\renewcommand{\H}{\mc{H}}
\newcommand{\B}{\mc{B}}
\newcommand{\hH}{\hat{H}}
\newcommand{\hA}{\hat{A}}
\newcommand{\hB}{\hat{B}}
\newcommand{\hS}{\hat{S}}
\newcommand{\dmp}{\rho}
\newcommand{\dm}{\hat\rho}
\newcommand{\opt}{{\operatorname{opt}}}
\newcommand{\aux}{{\operatorname{aux}}}
\newcommand{\trunc}{{\operatorname{trunc}}}
\newcommand{\odd}{{\operatorname{odd}}}
\newcommand{\even}{{\operatorname{even}}}
\newcommand{\supp}{{\operatorname{supp}}}
\renewcommand{\max}{{\operatorname{max}}}
\renewcommand{\vec}[1]{{\boldsymbol{#1}}}

\newcommand{\norm} [1]{\left\Vert #1\right\Vert}
\newcommand{\normS}[1]{\Vert #1\Vert}

\newcommand{\lmu} {Department of Physics and Arnold Sommerfeld Center for Theoretical Physics,
Ludwig-Maximilians-Universit{\"a}t M{\"u}nchen, Theresienstr.\ 37, 80333 Munich, Germany}
\newcommand{\Title} {Precise evaluation of thermal response functions by optimized density matrix renormalization group schemes}
\newcommand{\Authors}
{
\author{Thomas Barthel}
\affiliation{\lmu}
}
\newcommand{\Date} {December 20, 2012}

\begin{document}

\title{\Title}
\Authors

\begin{abstract}
This paper provides a study and discussion of earlier as well as novel more efficient schemes for the precise evaluation of finite-temperature response functions of strongly correlated quantum systems in the framework of the time-dependent density matrix renormalization group (tDMRG). The computational costs and bond dimensions as functions of time and temperature are examined for the example of the spin-$1/2$ XXZ Heisenberg chain in the critical XY phase and the gapped N\'{e}el phase. The matrix product state purifications occurring in the algorithms are in one-to-one relation with corresponding matrix product operators. This notational simplification elucidates implications of quasi-locality on the computational costs. Based on the observation that there is considerable freedom in designing efficient tDMRG schemes for the calculation of dynamical correlators at finite temperatures, a new class of optimizable schemes, as recently suggested in arXiv:1212.3570, is explained and analyzed numerically. A specific novel near-optimal scheme that requires no additional optimization reaches maximum times that are typically increased by a factor of two, when compared against earlier approaches. These increased reachable times make many more physical applications accessible. For each of the described tDMRG schemes, one can devise a corresponding transfer matrix renormalization group (TMRG) variant.
\end{abstract}

\date{\Date}


\maketitle
\begin{spacing}{0.4}
\tableofcontents
\end{spacing}

\newpage
\section{Introduction}
This paper addresses the efficient evaluation of finite-temperature response functions
\begin{equation}\label{eq:chiAB}
	\chi_{\hA\hB}(\beta,t):=\frac{1}{Z_\beta}\Tr(e^{-\beta \hH}\hB(t)\hA)\quad\text{with}\quad
	\hB(t)\equiv  e^{i\hH t}\hB e^{-i\hH t}
\end{equation}
for quantum many-particle systems on one-dimensional (1D) lattices in the framework of the \emph{density matrix renormalization group} (DMRG) \cite{White1992-11,White1993-10,Schollwoeck2005}. From the theoretical perspective, such quantities occur for example in the context of linear response theory \cite{Fetter1971,Negele1988} and characterize the effect of a perturbation $\hA$ of the system at time zero on the expectation value of an observable $\hB$ at time $t$. Initially, the system is in thermal equilibrium, where $\beta=1/T$ is the inverse temperature and $Z_\beta=\Tr e^{-\beta \hH}$. The units are chosen such that Boltzmann's and Planck's constants are $k_B=1$ and $\hbar=1$. Such response functions contain important information on the many-body physics and are addressed in many experimental setups. For example, recent advances in neutron-scattering techniques make very precise measurements possible. See for example Refs.\ \cite{Stone2003-91,Zaliznyak2004-93,Zaliznyak2005,Lake2005-4,Xu2007-317}. It is hence very important to have numerical tools at hand that allow for an efficient and precise evaluation of thermal response functions for (strongly-correlated) condensed matter models in order to match theoretical models to actual materials and to gain an understanding of the underlying physical processes. 

As discussed and demonstrated in several works \cite{Barthel2009-79b,Feiguin2010-81,Karrasch2012-108}, finite-temperature response functions for strongly-correlated 1D systems can be evaluated up to some maximum reachable time by using the time-dependent density matrix renormalization group (tDMRG) \cite{Vidal2003-10,White2004,Daley2004} applied to a \emph{purification} \cite{Uhlmann1976,Uhlmann1986,Nielsen2000} of the density matrix. In the DMRG approach, those purifications are approximated with a controllable precision by \emph{matrix product states} (MPS) \cite{Accardi1981,Fannes1991,Rommer1997}. To address finite-temperature states with tDMRG was suggested in Refs.~\cite{Verstraete2004-6,Zwolak2004-93} with first applications in Refs.~\cite{Feiguin2005-72,Barthel2005}. A difficulty in the simulations of time-evolved states is the growth of entanglement with time \cite{Calabrese2005,Bravyi2006-97,Eisert2006-97,Barthel2008-100}. In tDMRG calculations, this leads to a corresponding, typically exponential, increase of the computation cost with time and a strong limitation of the maximum reachable times, depending on the available computational resources and the desired accuracy of the simulation. The effect is much more drastic for mixed states. The focus of this paper is hence to study the computation cost of the different tDMRG schemes for the evaluation of thermal response functions and to suggest novel more efficient approaches that allow for a substantial increase in the maximum reachable times.

The results of such simulations can be Fourier transformed to study the spectral properties of the response. In order to avoid ringing artifacts from the Fourier transformation of the data on a restricted time interval, one can either use filters, which result in an artificial broadening, or use linear prediction \cite{Yule1927-226,Makhoul1975-63}. This was employed first in Ref.~\cite{White2008-77} for $T=0$ and in Ref.~\cite{Barthel2009-79b} for $T>0$. See also Refs.~\cite{Pereira2012-85,Ren2012-85}, for further applications of linear prediction in DMRG calculations at $T=0$.

Earlier alternative DMRG approaches for finite-temperature response functions have built on the \emph{transfer matrix renormalization group} (TMRG) \cite{Nishino1995-64,Bursill1996-8,Shibata1997-66,Wang1997-56}. In a first step, analytic continuation techniques as in \emph{Quantum Monte Carlo} (QMC) were employed \cite{Naef1999-60}. In a second development, transfer matrices comprising the imaginary- and the real-time evolution were used to access autocorrelation functions \cite{Sirker2005-71,Sirker2006-73}. In finite-temperature QMC calculations (e.g.\ positive-definite path integral \cite{Suzuki1977-58,Hirsch1982-26} or stochastic series expansion with directed loops \cite{Sandvik1991-43}), real-time or real-frequency response functions have to be extracted by analytic continuation from imaginary-time results \cite{Jarrell1996-269} which is ill-conditioned and numerically challenging. Finite temperatures have also been addressed by a hybrid algorithm based on Monte Carlo and tDMRG \cite{White2009-102,Stoudenmire2010-12}.

In this paper, I discuss how the simulation based on the evolution of a certain MPS purification of the thermal density matrix as described in \cite{Verstraete2004-6,Barthel2009-79b,Karrasch2012-108} can actually be understood as evolving a \emph{matrix product operator} (MPO) representation of the square root of the thermal density matrix. The description in terms of purifications is in this sense superfluous. A decisive observation is now that there are considerable degrees of freedom for the choice of a specific evaluation scheme. For the two specific schemes, corresponding to Refs.~\cite{Barthel2009-79b,Feiguin2010-81} (\emph{scheme A}) and Ref.~\cite{Karrasch2012-108} (\emph{scheme B}), respectively, I examine the scaling of the computation cost with time. It turns out that \emph{scheme A} has some advantage at low temperatures and \emph{scheme B} is advantageous at higher temperatures, for which I give an explanation on the basis of quasi-locality \cite{Nachtergaele2007-12a,Barthel2012-108b}. The maximum reachable times $t^{A}_\max(\beta)$ and $t^{B}_\max(\beta)$ differ, for the same computation cost and accuracy, by a factor of order one. It is then shown that an optimized evaluation scheme, for which one optimizes over the aforementioned degrees of freedom, yields maximum reachable times that are at least twice as large as for \emph{scheme B}, specifically $t^\opt_\max(2\beta)\gtrsim 2 t^B_\max(\beta)$. In all cases one finds that $t_\max(2\beta)\approx t_\max(\beta)$ as $t_\max$ varies slowly as a function of $\log\beta$. We can devise a new \emph{scheme C} that does not require any optimization, but outperforms \emph{scheme B} by a factor of two in the maximum reachable time. It is only outdone by \emph{scheme A} at very low temperatures. To go beyond \emph{scheme C}, one can study the computation cost for calculating MPO approximations of operators $e^{i\hH s}e^{-\alpha\hH}\hB e^{-i\hH s'}$ to determine optimized evolution schemes. Finally, the connection between \emph{scheme A} and an earlier TMRG approach \cite{Sirker2005-71} is explained. For each of the tDMRG schemes, one can devise a corresponding TMRG variant.

Due to the substantially increased reachable times and its simplicity (no need for optimization), the novel \emph{scheme C}, which was introduced recently in Ref.~\cite{Barthel2012_12unused} and is explained here in detail, can be expected to be the method of choice for future applications. In Ref.~\cite{Barthel2012_12unused}, it allowed us to show that the thermal spectral functions of 1D bosons in the quantum critical regime with dynamic critical exponent $z=2$ follows a universal scaling form and to compute the corresponding scaling function precisely.

\section{MPS purification versus MPO density matrix}\label{sec:DMRG}
As described for example in Refs.~\cite{Verstraete2004-6,Barthel2009-79b,Karrasch2012-108}, the (unnormalized) thermal density matrix
\begin{equation*}
	\dm_\beta:=e^{-\beta \hH},\quad Z_\beta:=\Tr e^{-\beta \hH}
\end{equation*}
for a chain of $L$ sites can be obtained in the form of an MPS purification
\begin{subequations}\label{eq:mps_purify}
\begin{gather}
	|\dmp_\beta\ket \in \H\otimes\H_\aux,\quad \H_\aux\simeq \H,\\
	|\dmp_\beta\ket = \sum_{\vec{\sigma},\vec{\sigma}'} A^{\sigma_1, \sigma'_1}_1 A^{\sigma_2, \sigma'_2}_2\dotsm A^{\sigma_L, \sigma'_L}_L
	 \underbrace{|\sigma_1\sigma_2\dots \sigma_L\ket}_{=:|\vec{\sigma}\ket}\otimes \underbrace{|\sigma'_1 \sigma'_2\dots \sigma'_L\ket_\aux}_{=:|\vec{\sigma}'\ket_\aux},
\end{gather}
\end{subequations}
where the $|\sigma_i\ket$ label orthonormal site basis states for $\H$, the auxiliary Hilbert space $\H_\aux$ is isomorphic to $\H$, and $|\sigma'_i\ket_\aux$ label orthonormal site basis states for $\H_\aux$. The MPS is constructed from $M_{i-1}\times M_i$ matrices $A^{\sigma_i,\sigma'_i}_i$, where $M_0=M_L=1$. The matrix sizes $\{M_i\}$ are also called \emph{bond dimensions}; see for example the review \cite{Schollwoeck2005}.

In order for $|\dmp_\beta\ket$ to be a purification of $\dm_\beta$, it needs to be constructed such that
\begin{equation*}
	\Tr_\aux|\dmp_\beta\ket\bra\dmp_\beta|
	\equiv \sum_{\vec{\sigma}\vec{\sigma}'\vec{\sigma}''}|\vec{\sigma}\ket\bra \vec{\sigma}\vec{\sigma}'|\dmp_\beta\ket\bra\dmp_\beta|\vec{\sigma}''\vec{\sigma}'\ket\bra \vec{\sigma}''|
	=\dm_\beta.
\end{equation*}
This can be achieved by choosing the infinite-temperature purification to be
\begin{equation*}
	|\dmp_0\ket=|\id\ket\quad\text{with}\quad
	|\id\ket := \sum_{\vec{\sigma}}|\vec{\sigma}\ket\otimes|\vec{\sigma}\ket_\aux
	 = \bigotimes_{i=1}^L\Big(\sum_{\sigma_i} |\sigma_i\ket\otimes|\sigma_i\ket_\aux\Big),
\end{equation*}
which can be written as an MPS \eqref{eq:mps_purify} with bond dimensions $M_i=1$.
In this so-called ancilla approach, finite-temperature purifications can be calculated by imaginary-time evolution
\begin{equation*}
	|\dmp_\beta\ket =  e^{-\beta \hH/2}\otimes \id_\aux\cdot |\id\ket.
\end{equation*}
and response functions are evaluated after a subsequent real-time evolution
\begin{equation}\label{eq:chi_purify}
	\chi_{\hA\hB}(\beta,t) = \frac{1}{Z_\beta}\big[\bra\dmp_\beta| e^{i\hH t}\big]\hB \big[e^{-i\hH t}\hA |\dmp_\beta\ket\big].
\end{equation}
The evolution of the MPS can for example be implemented by decomposing the propagators into circuits of local gates by a Trotter-Suzuki decomposition \cite{Vidal2003-10,White2004,Daley2004} as described in Section~\ref{sec:tDMRG_vs_TMRG}. The square brackets in Eq.~\eqref{eq:chi_purify} indicate which parts of the expression are represented as MPS after the evolution.
Note that $Z_\beta$, which is required in Eq.~\eqref{eq:chi_purify}, can be determined as $\bra\dmp_\beta|\dmp_\beta\ket=\Tr e^{-\beta\hH} =Z_\beta$. Equivalently, one can work with purifications $|\dmp_\beta\ket$ that are normalized to one.

In each step of the tDMRG, the evolved states $|\psi\ket = |\psi(\beta,t)\ket$ are approximated by an MPS with bond dimensions $M_i=M_i(\beta,t)$ that are as small as possible for a given constraint on the desired precision of the approximation \cite{Schollwoeck2005}. This is achieved through \emph{truncations}. For every splitting of the system into a left and a right part, one does a Schmidt decomposition $|\psi\ket=\sum_{k=1}^{\tilde M}\lambda_k|k\ket_L\otimes|k\ket_R$ of the state \cite{Nielsen2000} which boils down to doing singular value decompositions of the tensors $A_i$. The corresponding reduced density matrices are $\sum_k\lambda_k^2|k\ket_L\bra k|_L$ and $\sum_k\lambda_k^2|k\ket_R\bra k|_R$. The bond dimension is then reduced from $\tilde M$ to some value $M<\tilde M$ by retaining only the $M$ largest Schmidt coefficients and truncating all smaller ones.
\begin{equation}
	 |\psi\ket=\sum_{k=1}^{\tilde M}\lambda_k|k\ket_L\otimes|k\ket_R\quad\mapsto\quad
	 |\psi_\trunc\ket=\sum_{k=1}^{M}\lambda_k|k\ket_L\otimes|k\ket_R
\end{equation}
The precision is in each step of the algorithm controlled by bounding the \emph{truncation weight}
\begin{equation}
	\epsilon=\left(\frac{\normS{\psi_\trunc - \psi}}{\normS{\psi}}\right)^2 = \frac{\sum_{k>M}\lambda_k^2}{\sum_k\lambda_k^2}.
\end{equation}

\begin{figure}[t]
\includegraphics[width=0.9\textwidth]{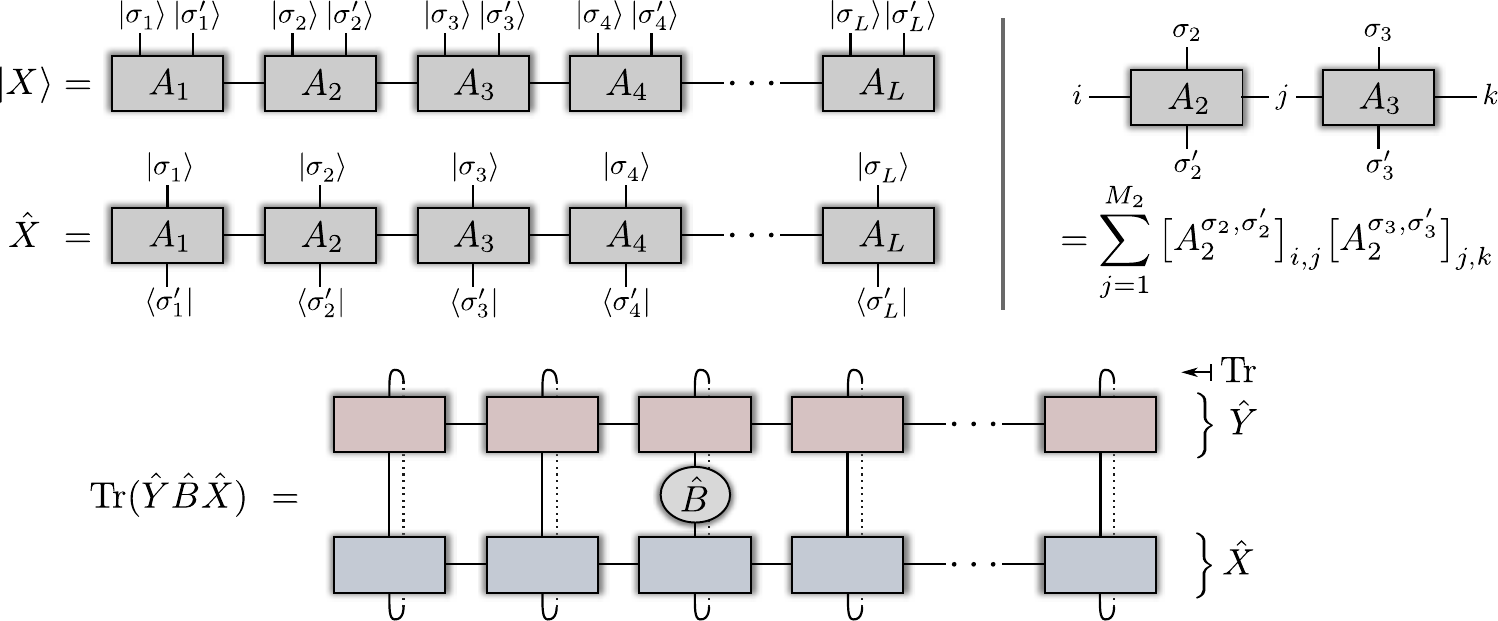}
\caption{\label{fig:mpo}An MPS purification $|X\ket$ and the corresponding MPO $\hat X$, according to the isomorphism \eqref{eq:mpIsomorphism}. In the graphical representation, boxes represent the tensors $A_i$ that define the MPS or MPO, and lines represent partial tensor contractions. The diagram for $\Tr(\hat Y\hat B\hat X)$ represents the tensor network that has to be contracted in order to evaluate the trace of two MPOs $\hat X$ and $\hat Y$, and a local observable $\hat B$ as occurring in the tDMRG schemes for the evaluation of thermal response functions. See for example Eq.~\eqref{eq:chi_schemeA}. The computational cost is determined by the bond dimensions of the MPOs as in Eq.~\eqref{eq:cost}.}
\end{figure}
As a matter of fact, the notation of this procedure as an algorithm on purifications is in a sense superfluous. Due to the fact that $\B(\H)$, the space of linear maps on $\H$, and the tensor product space $\H\otimes\H$ are isomorphic, everything can be rewritten in terms of MPOs in $\B(\H)$ instead of MPS purifications in $\H\otimes\H$,
\begin{equation}\label{eq:mpIsomorphism}
	\bra\vec{\sigma}\vec{\sigma}'|X\ket \equiv \bra\vec{\sigma}|\hat X|\vec{\sigma}'\ket\quad\forall_{\vec{\sigma}\vec{\sigma}'}.
\end{equation}
Examples for this one-to-one relation are
\begin{alignat*}{4}
	|\id\ket \quad&\longleftrightarrow&&\quad \id,\\
	|\dmp_\beta\ket \quad&\longleftrightarrow&&\quad e^{-\beta\hH/2},\\
	\hat Y\otimes\hat Z_\aux |X\ket \quad&\longleftrightarrow&&\quad \hat Y\hat X \hat Z^T,
\end{alignat*}
where the transposition is to be executed in the $|\vec{\sigma}\ket$ basis.
In particular, the counterpart of the MPS purification \eqref{eq:mps_purify} is the MPO
\begin{equation}\label{eq:mpo}
	\sum_{\vec{\sigma}\vec{\sigma}'} A^{\sigma_1, \sigma'_1}_1 A^{\sigma_2, \sigma'_2}_2\dotsm A^{\sigma_L, \sigma'_L}_L
	 |\vec{\sigma}\ket\bra\vec{\sigma}'|.
\end{equation}
See for example Refs.~\cite{Zwolak2004-93,McCulloch2007-10,Schollwoeck2011-326} for discussions of MPOs.

Applying evolution operators $e^{-\Delta\beta \hH}$ or $e^{\pm i\Delta t \hH}$ to the left or right of an MPO $\hat X$ can be done by tDMRG in exactly the same way as for MPS, for example, by a Trotter-Suzuki decomposition as described in Section~\ref{sec:tDMRG_vs_TMRG}. The truncation weight, which is kept below a certain bound in order to control the precision, is then given by
\begin{equation}\label{eq:MPOtrunc}
	\epsilon=\left(\frac{\normS{\hat X_\trunc - \hat X}_2}{\normS{\hat X}_2}\right)^2,
\end{equation}
where $\normS{\hat X}_{2}=\sqrt{\Tr \hat X^\dag \hat X}$ is the Schatten 2-norm. This is the natural norm within the DMRG framework.

\section{Evaluation \emph{schemes A} and \emph{B}}\label{sec:schemesAB}
\begin{figure}[t]
\includegraphics[width=0.8\textwidth]{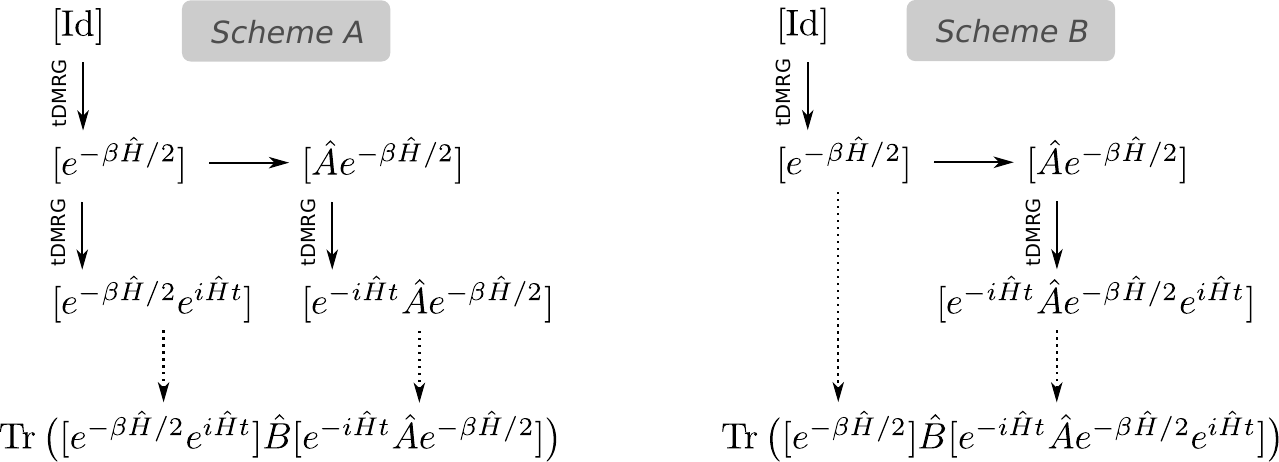}
\caption{\label{fig:schemesAB}\emph{Schemes A} and \emph{B} for the evaluation of the response function according to equations \eqref{eq:chi_schemeA} and \eqref{eq:chi_schemeB}. In the tDMRG simulations, the operators in square brackets are approximated as MPOs. The computation costs \eqref{eq:cost} for given $t$ and $\beta$ depend on the chosen scheme and the desired accuracy \eqref{eq:MPOtrunc} of the MPO approximations.}
\end{figure}

Evaluation \emph{scheme A} as described and employed in Refs.~\cite{Barthel2009-79b,Feiguin2010-81} corresponds to Eq.~\eqref{eq:chi_purify}. In the MPO notation it reads
\begin{equation}\label{eq:chi_schemeA}
	\chi^A_{\hA\hB}(\beta,t) = \frac{1}{Z_\beta}\Tr\left(\big[e^{-\beta\hH/2} e^{i\hH t}\big]\hB \big[e^{-i\hH t}\hA e^{-\beta\hH/2}\big]\right).
\end{equation}
The square brackets indicate which parts of the expression are represented as MPOs after the evolution. Figure~\ref{fig:schemesAB} shows the algorithm diagrammatically. The partition function $Z_\beta$ can be obtained from the Schatten 2-norm $Z_\beta=(\normS{[e^{-\beta\hH/2}]}_{2})^2$. 
In the DMRG method, multiplications of the matrices $A_i^{\sigma_i, \sigma'_i}$ and singular value decompositions, required for the truncations of the MPOs, dominate the computation costs. Hence, the computation costs for a single evolution step for the first MPO (bond dimensions $M_i$) and the second MPO (bond dimensions $M_i'$), and for the final evaluation of Eq.~\eqref{eq:chi_schemeA} scale as
\begin{equation}\label{eq:cost}
	\sum_{i=1}^L M_i^3,\quad \sum_{i=1}^L (M'_i)^3,\quad\text{and}\quad \sum_{i=1}^L \max\left(M_i(M'_i)^2,M_i^2 M'_i\right),
\end{equation}
respectively \cite{foot:HartleyEntropy}\nocite{Hartley1928-7,Klir2005}; see Figure~\ref{fig:mpo}.

For a nonsingular operator $\hat T\in\B(\H)$, the value of \eqref{eq:chi_schemeA} is invariant under transformations
\begin{subequations}\label{eq:trafo}
\begin{alignat}{4}
	\big[e^{-\beta\hH/2} e^{i\hH t}\big]
	&\quad\longrightarrow\quad&& \big[\hat T e^{-\beta\hH/2} e^{i\hH t}\big]\\
	\big[e^{-i\hH t}\hA e^{-\beta\hH/2}\big]
	&\quad\longrightarrow\quad&& \big[e^{-i\hH t}\hA e^{-\beta\hH/2}\hat T^{-1}\big]
\end{alignat}
\end{subequations}
of the involved MPOs \cite{foot:commute}. This is a considerable degree of freedom that can be exploited to reduce, for given $\beta$, $t$, and $\epsilon$, the bond dimensions $M_i$ of the MPOs and, hence, to reduce the computation cost. In cases where such a reduction is possible, one can extend the simulation to longer times $t$. Some numerical experiments showed that the corresponding computation cost for the optimization of $\hat T$ scales exponentially with the system size. So a search for globally optimal $\hat T$ seems to be a hopeless exercise. It is hence reasonable to constrain ourselves to certain classes of transformations, e.g., $\hat T=e^{-\beta'\hH}e^{-i\hH t'}$ and to optimize with respect to the degrees of freedom of the class -- $\beta'$ and $t'$ in that case. See Section~\ref{sec:opt}.

The modified evaluation \emph{scheme B} (see Fig.~\ref{fig:schemesAB}) employed in Refs.~\cite{Karrasch2012-108,Huang2012_12} corresponds to the choice $\hat T=e^{-i\hH t}$ and reads in the operator notation
\begin{align}
	\chi^B_{\hA\hB}(\beta,t) &= \frac{1}{Z_\beta}\Tr\left(\big[e^{-i\hH t}e^{-\beta\hH/2} e^{i\hH t}\big]\hB \big[e^{-i\hH t}\hA e^{-\beta\hH/2}e^{i\hH t}\big]\right)\nonumber\\
		 \label{eq:chi_schemeB}
	&= \frac{1}{Z_\beta}\Tr\left(\big[e^{-\beta\hH/2}\big]\hB \big[e^{-i\hH t}\hA e^{-\beta\hH/2}e^{i\hH t}\big]\right).
\end{align}

\section{Costs of \emph{schemes A} and \emph{B} and their explanation}
\begin{figure}[t]
\includegraphics[width=0.55\textwidth]{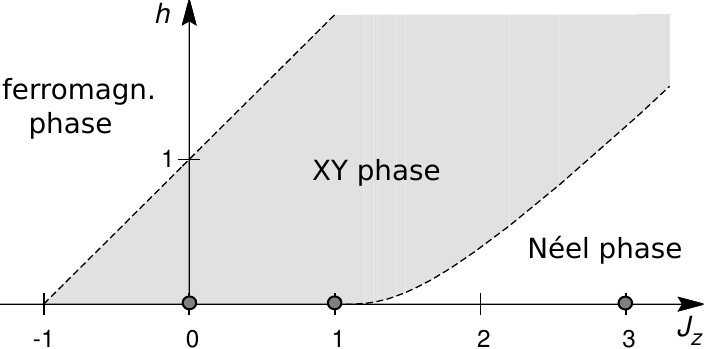}
\caption{\label{fig:XXZ}Zero-temperature phase diagram of the spin-$1/2$ XXZ Heisenberg model in a magnetic field \eqref{eq:HamXXZ}. The ground state is fully polarized in the ferromagnetic phase. In the N\'{e}el phase, it has a finite staggered magnetization. The system is critical (vanishing energy for excitations) in the XY, or spin-liquid, phase (gray). The phase boundaries can be obtained from the Bethe ansatz \cite{Cloizeaux1966-7,Johnson1972-6}. In this work, the properties of different tDMRG schemes for the evaluation of thermal response functions are exemplified with this model at the three points $J_z=0,1,3$  with $h=0$.}
\end{figure}
\begin{figure}[p]
\includegraphics[width=0.95\textwidth]{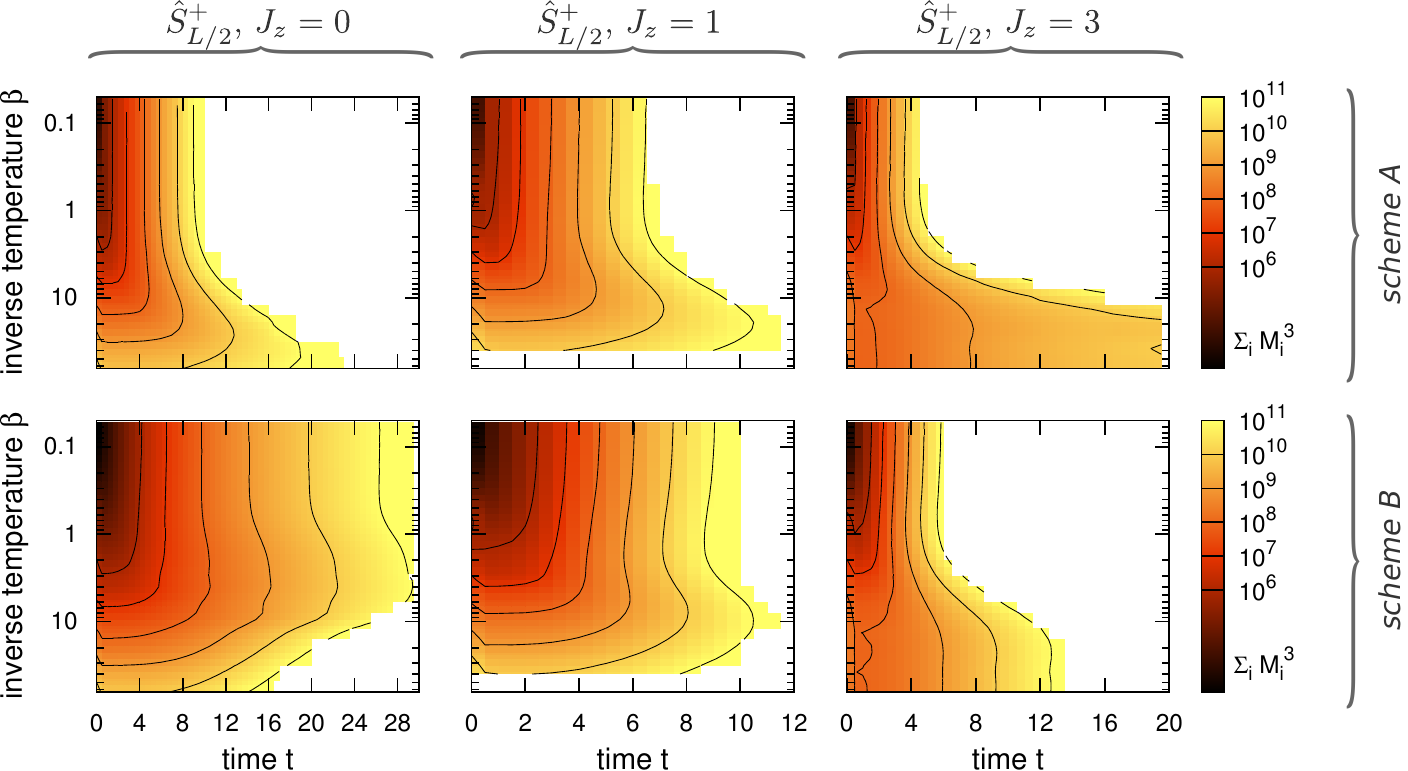}
\caption{\label{fig:AB_Sp63_cost}Computation cost per time step ($\sum_i \left(M_i(\beta,t)\right)^3$) for \emph{scheme A} (top) and \emph{scheme B} (bottom) with $J_z=0,1,$ and $3$, $\hA=\hB^\dag=\hS^+_{L/2}$ \cite{foot:A_equal_Bh}, and truncation weights per time step $\epsilon_\beta=10^{-12}$, $\epsilon_t=10^{-10}$.
The contour lines correspond to maximum reachable times for different computational resources (per time step) and the predefined precision of the simulations ($\epsilon_\beta$, $\epsilon_t$).
}
\end{figure}
\begin{figure}[p]
\includegraphics[width=0.95\textwidth]{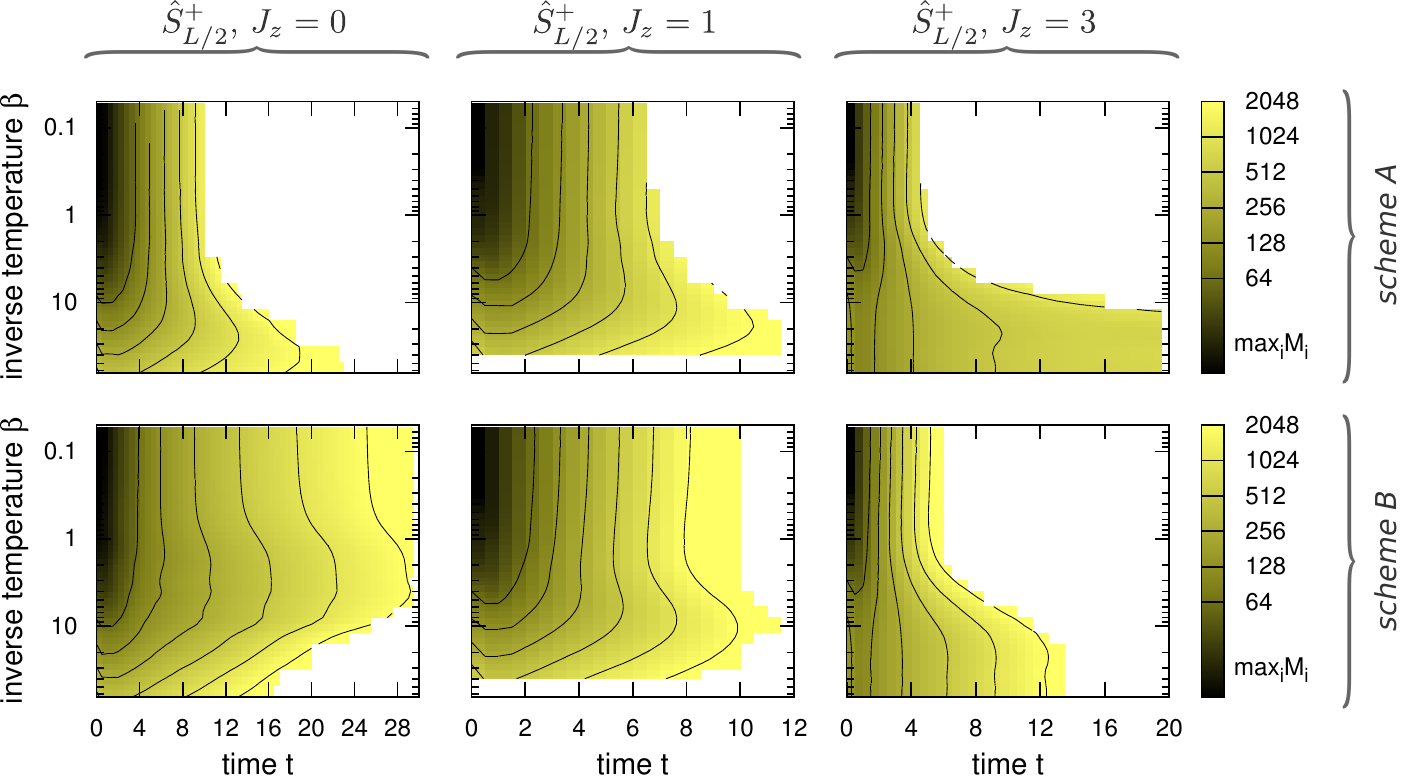}
\caption{\label{fig:AB_Sp63_dim}Maximum bond dimensions $\max_i M_i(\beta,t)$ for \emph{scheme A} (top) and \emph{scheme B} (bottom) with $J_z=0,1,$ and $3$, $\hA=\hB^\dag=\hS^+_{L/2}$, and truncation weights per time step $\epsilon_\beta=10^{-12}$, $\epsilon_t=10^{-10}$.}
\end{figure}
In order to study the computational costs of evaluation \emph{schemes A} and \emph{B} [Eqs.~\eqref{eq:chi_schemeA} and \eqref{eq:chi_schemeB}], as functions of time and temperature, let us choose as an example the spin-$1/2$ XXZ Heisenberg model
\begin{equation}\label{eq:HamXXZ}
	 \hat H = \sum_{i=1}^{L-1}(\hS^x_i\hS^x_{i+1}+\hS^y_i\hS^y_{i+1}+J_z \hS^z_i\hS^z_{i+1}) -h\sum_{i=1}^L\hS^z_i.
\end{equation}
The phase diagram \cite{Cloizeaux1966-7,Johnson1972-6,Mikeska2004}, as derived by the Bethe ansatz, is shown in Figure~\ref{fig:XXZ}.
The simulations are carried out for vanishing magnetic field $h=0$, system size $L=128$, and coupling constants $J_z=0$ and $J_z=1$, where the model is gapless, and at $J_z=3$, where the model is in its gapped antiferromagnetic phase. For the time evolution, a fourth order Trotter-Suzuki decomposition with step sizes $\Delta \beta=\Delta t=1/8$ is used. The truncation weights in the imaginary-time and real-time evolutions were fixed to values $\epsilon_\beta$ and $\epsilon_t$ which are specified in the captions of the corresponding figures.

Figure~\ref{fig:AB_Sp63_cost} shows the computation cost per time step for the operators $\hA=\hB^\dag=\hS^+_{L/2}$ in Eqs.~\eqref{eq:chi_schemeA} and \eqref{eq:chi_schemeB}, i.e., spin-flips in the middle of the system \cite{foot:A_equal_Bh}. The density plots show, as a function of time and temperature, the number of operations needed for a single time step as measured by $\sum_i \left(M_i(\beta,t)\right)^3$. For \emph{scheme A}, the cost is dominated by the cost for the computation of the MPO
\begin{equation}\label{eq:costlyOp_A}
	\big[e^{-i\hH t}\hA e^{-\beta\hH/2}\big]. 
\end{equation}
For \emph{scheme B}, the computation of the MPO
\begin{equation}\label{eq:costlyOp_B}
	\big[e^{-i\hH t}\hA e^{-\beta\hH/2}e^{i\hH t}\big]
\end{equation}
dominates the numerical costs. Hence, the bond dimensions $M_i(\beta,t)$ of those operators were used in the diagrams.
Please notice the logarithmic scale for $\beta=1/T$ and the cost, i.e., the color coding and the distances of the contour lines \cite{foot:HartleyEntropy}. The contour lines correspond to maximum reachable times for certain computational resources and a predefined precision of the simulation, determined by $\epsilon_\beta$ and $\epsilon_t$. Figure~\ref{fig:AB_Sp63_dim} shows, for the same simulations, the maximum bond dimensions $\max_{i\in[1,L]}M_i(\beta,t)$.

\begin{figure}[p]
\includegraphics[width=0.985\textwidth]{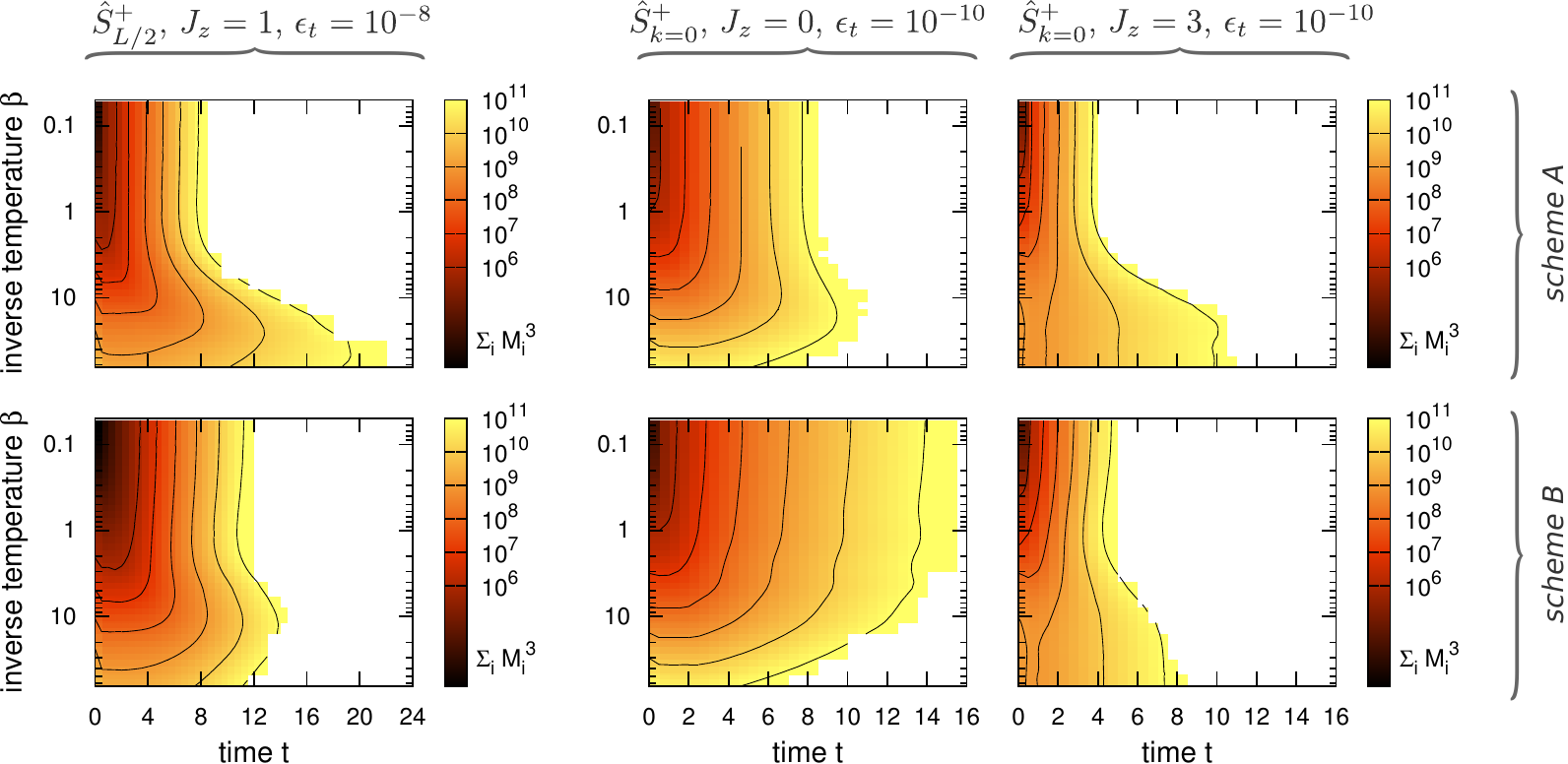}
\caption{\label{fig:AB_Sp63Sk0_cost}Computation cost per time step ($\sum_i \left(M_i(\beta,t)\right)^3$) for \emph{schemes A} and \emph{B}. Left: $J_z=1$, and $\hA=\hB^\dag=\hS^+_{L/2}$ with (increased) truncation weights $\epsilon_\beta=10^{-10}$, $\epsilon_t=10^{-8}$. Right: $J_z=1$ and $3$, and $\hA=\hB^\dag=\hS^+_{k=0}$ with the usual truncation weights $\epsilon_\beta=10^{-12}$, $\epsilon_t=10^{-10}$.}
\end{figure}

First of all, the results show that the computation costs increase, for fixed temperature, exponentially with time $t$. This corresponds to a linear increase of the entanglement entropy in the evolution of pure states after a quench \cite{Calabrese2005,Bravyi2006-97,Eisert2006-97}. For the noncritical case, $J_z=3$, the costs per time step become $\beta$-independent for temperatures that are sufficiently below the excitation gap.
At higher temperatures, the cost for \emph{scheme B} is systematically smaller than that for \emph{scheme A}. The effect is strongest for the non-interacting system at $J_z=0$. For $J_z=1$ and $J_z=3$, the increase in the maximum reachable times is more moderate with a factor of $\lesssim 1.4$. At lower temperatures, \emph{scheme A} is more efficient than \emph{scheme B}. This trend strengthens when the accuracy of the simulation is reduced (higher truncation weights $\epsilon_\beta$ and $\epsilon_t$) as documented in the left part of Figure~\ref{fig:AB_Sp63Sk0_cost}.
Figure~\ref{fig:AB_Sp63_dim} shows that the maximum bond dimensions $\max_i M_i(\beta,t)$ evolve almost in the same way as the computation costs. For the noncritical case $J_z=3$, one sees however that the maximum bond dimensions occurring in \emph{scheme A} are for all temperatures smaller than those occurring in \emph{scheme B}, whereas the computation cost for \emph{scheme B} is lower at higher temperatures. Also, for $J_z=1$, the maximum bond dimensions occurring in the two schemes are quite similar.
\begin{figure}[p]
\includegraphics[width=0.985\textwidth]{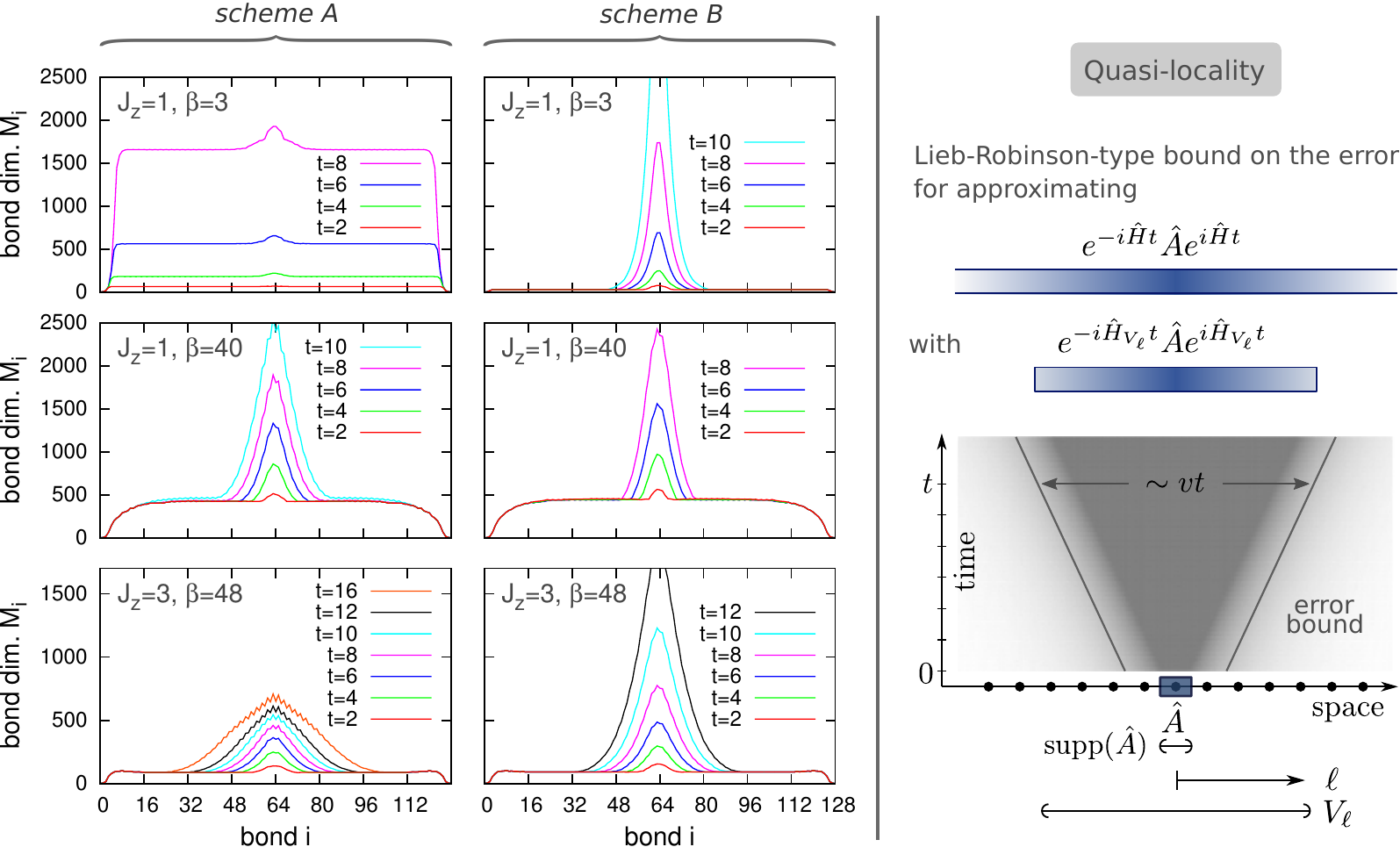}
\caption{\label{fig:AB_Sp63_slices}Evolution of the MPOs \eqref{eq:costlyOp_A} and \eqref{eq:costlyOp_B} as occurring in \emph{schemes A} and \emph{B}, respectively. The plots show the bond dimensions $M_i=M_i(\beta,t)$ for $\hA=\hS^+_{L/2}$, high as well as low temperatures, and the critical point $J_z=1$ as well as the gapped $J_z=3$. The truncation weights were $\epsilon_\beta=10^{-12}$ and $\epsilon_t=10^{-10}$. Unlike in \emph{scheme A}, the MPO for \emph{scheme B} remains unchanged outside a certain space-time cone also at high temperatures. This is due to the quasi-locality (right); see Eq.~\eqref{eq:quasilocal}.}
\end{figure}

These findings can be explained as follows. In Ref.~\cite{Karrasch2012-108} it was pointed out that the operator \eqref{eq:costlyOp_B} of \emph{scheme B} is time-independent for the simple case $\hA=\id$, whereas the computation cost for the MPO \eqref{eq:costlyOp_A}, occurring in \emph{scheme A}, can increase with time, even in this trivial case. More generally, the following argument applies for all operators $\hA$ with finite spatial support $\supp(\hA)$. The typical condensed matter systems are \emph{quasi-local} \cite{Nachtergaele2007-12a,Barthel2012-108b}, i.e., the spatial support of operators like $e^{-i\hH t}\hA e^{i\hH t}$, occurring in \emph{scheme B}, grows only linearly with time. More precisely, outside a certain space-time cone originating from $\supp(\hA)$, the evolved operator acts almost like the identity and does hence not change the entanglement in that region; see Figure~\ref{fig:AB_Sp63_slices}. In mathematical terms, quasi-locality means
\begin{equation}\label{eq:quasilocal}
	\norm{e^{-i\hH t}\hA e^{i\hH t}-e^{-i\hH_V t}\hA e^{i\hH_V t}}
	 \leq C\normS{\hA} e^{vt-\operatorname{dist}\left(\partial V,\supp(\hA)\right)},
\end{equation}
where all terms $\hat h_i$ in the Hamiltonian $\hH=\sum_i\hat h_i$ are required to be short-ranged and norm-bounded, $\hH_V=\sum_{i\in V}\hat h_i$ is the Hamiltonian truncated to a vicinity $V$ of the spatial support of the operator $\hA$ ($\supp(\hA)\subset V$), $\partial V$ is the boundary of $V$, and $v\propto \operatorname{sup}_i \normS{\hat h_i}$ is the Lieb-Robinson velocity \cite{Nachtergaele2007-12a,Barthel2012-108b}. So the norm-distance between the exactly evolved operator and the operator, evolved on the subsystem $V$ only, decays exponentially with increasing distance of $\partial V$ to a space-time cone defined by the Lieb-Robinson velocity $v$, and originating from $\supp(\hA)$ at time zero. As the evolved operator $e^{-i\hH t}\hA e^{i\hH t}$, hence, behaves up to exponentially small corrections like the identity outside the specified space-time cone, it does not alter the operator $\big[e^{-\beta\hH/2}\big]$ in that region -- in particular, not the MPO bond dimensions $M_i$ -- for the product \eqref{eq:costlyOp_B} occurring in \emph{scheme B}. This effect is very well visible in the plots for $J_z=1$ and $\beta=3$ shown in Figure~\ref{fig:AB_Sp63_slices}. For both tDMRG schemes, the maximum bond dimensions occur in the middle of the system, where $\hA=\hS^+_{L/2}$ acts, and grow (exponentially) with time $t$. The maxima have comparable values. Whereas, in \emph{scheme B}, $M_i$ remains unchanged outside the Lieb-Robinson space-time cone due to the quasi-locality, they grow for all bonds in \emph{scheme A}. This implies according to Eq.~\eqref{eq:cost} an increased cost for \emph{scheme A} at those temperatures.

Nevertheless, as Figures~\ref{fig:AB_Sp63_cost}--\ref{fig:AB_Sp63_slices} exemplify, \emph{scheme A} is often advantageous at low temperatures, especially for noncritical systems. For low temperatures, the quasi-locality is not essential as $e^{-\beta\hH/2}$ limits the effect of the real-time propagators $e^{\pm i\hH t}$ to a low-energy subspace. Sufficiently far away from the support of $\hA$, the resulting  dynamics is then very restricted and can not lead to a big increase of the bond dimensions. This is true for both schemes. Furthermore, in the middle of the system, where $\hA=\hS^+_{L/2}$ acts, the growth of the bond dimensions is slower for \emph{scheme A} which can be attributed to the fact that one acts with one propagator in Eq.~\eqref{eq:costlyOp_A} instead of acting with two propagators in Eq.~\eqref{eq:costlyOp_B}. As the plots for $J_z=1,3$ and $\beta=40,48$ in Figure~\ref{fig:AB_Sp63_slices} exemplify, the effect is much more pronounced for gapped systems.

The described properties have also been found for other local operators $\hA$ and $\hB$ like $\hS^z_{L/2}$ etc. The behavior for non-local operators like $\hS^+_{k=0}:=\sum_i\hS^+_i$ is also quite similar as exemplified in Figure~\ref{fig:AB_Sp63Sk0_cost} for $J_z=0,3$. Please note that $[\hS^+_{k=0},\hH]=0$ in the isotropic case $J_z=1$ for which the evolution of $\hS^+_{k=0}(t)$ is hence trivial.

\section{Optimized schemes and near-optimal \emph{scheme C}}\label{sec:opt}
\begin{figure}
\includegraphics[width=0.95\textwidth]{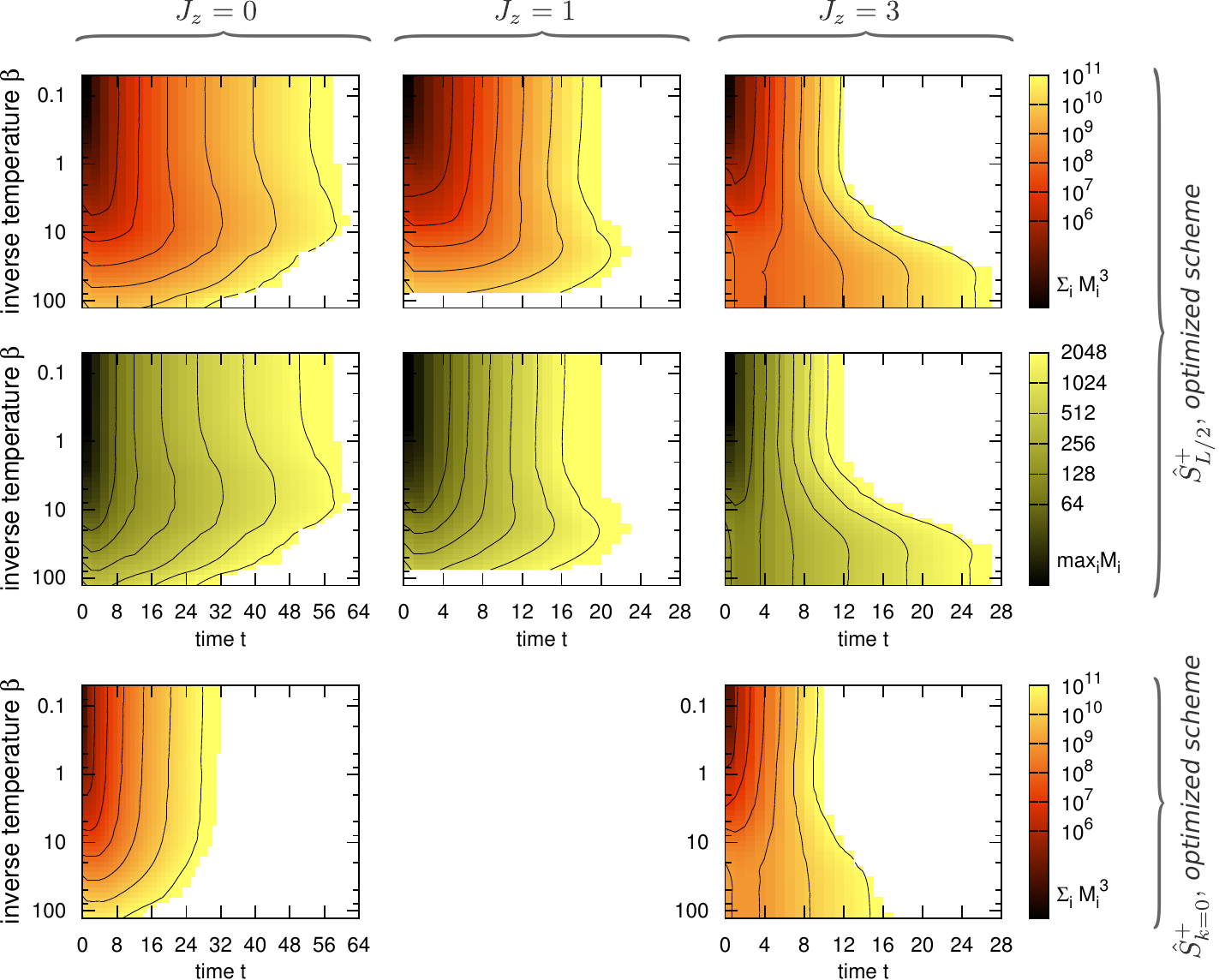}
\caption{\label{fig:opt_Sp63Spk0}The computation cost per time step $\sum_i \left(M_i(\beta,t)\right)^3$ (top and bottom) and the maximum bond dimension $\max_i M_i(\beta,t)$ (middle) for the optimized evaluation scheme \eqref{eq:chi_schemeFull2} with $J_z=0,1,$ and $3$, respectively. with $\hA=\hB^\dag=\hS^+_{L/2},\hS^+_{k=0}$, the schemes have been optimized with respect to $\beta'$ and $t'$. The corresponding efficiency measures were chosen to be the computation cost per time step for the top and bottom rows of diagrams, and the maximum bond dimension was chosen for the middle row of diagrams. In all computations, the truncation weights were chosen as $\epsilon_\beta=10^{-12}$, $\epsilon_t=10^{-10}$.}
\end{figure}
\emph{Schemes A} and \emph{B}, Eqs.~\eqref{eq:chi_schemeA} and \eqref{eq:chi_schemeB}, are typically far from optimal. One has a lot of freedom in designing a scheme that is as efficient as possible. The choice for a specific measure for the efficiency of a scheme can depend on the available computational resources. In this paper, the computation cost per time step as quantified by $\sum_i \left(M_i(\beta,t)\right)^3$ [Eq.~\eqref{eq:cost}] and, as an alternative, the maximum bond dimension $\max_i M_i(\beta,t)$ are considered as useful measures. With a nonsingular operator $\hat T$, the most generic scheme involving two MPOs is $Z_\beta^{-1}\Tr\big(\big[e^{i\hH t}\hB\hat T\big] \big[\hat T^{-1}e^{-i\hH t}\hA e^{-\beta\hH}\big]\big)$. As pointed out in Section~\ref{sec:schemesAB}, to optimize efficiently with respect to generic $\hat T$ is in general not possible. Thus, it is reasonable to constrain ourselves to a certain subclass of schemes and to optimize the efficiency over the remaining degrees of freedom of that class. Let us consider the class of schemes
\begin{equation}\label{eq:chi_schemeFull}
	\chi^{\beta't't''}_{\hA\hB}(\beta,t)
	= \frac{1}{Z_\beta}\Tr\left(\big[e^{i\hH t'}e^{-\beta'\hH}\hB e^{-i\hH t''}\big] \big[e^{-i\hH (t-t'')}\hA e^{-(\beta-\beta')\hH}e^{i\hH(t-t')}\big]\right).
\end{equation}
As before, the two factors in square brackets are to be approximated by MPOs \cite{foot:commute}. For given inverse temperature $\beta$, time $t$, and accuracy as quantified by the truncation weight $\epsilon$ [Eq.~\eqref{eq:MPOtrunc}] one can now optimize the chosen efficiency measure with respect to $\beta'$, $t'$, and $t''$. \emph{Scheme A} corresponds to the choice $(\beta',t',t'')=(\beta/2,t,0)$, \emph{scheme B} to $(\beta',t',t'')=(\beta/2,0,0)$, and the choice $(\beta',t',t'')=(\beta,0,0)$ for example to the Heisenberg picture. In actual applications, one can first study the computation cost, for obtaining operators $\big[e^{i\hH s}e^{-\alpha\hH}\hB e^{-i\hH s'}\big]$ and $\big[e^{-i\hH s}\hA e^{-\alpha\hH}e^{i\hH s'}\big]$. From the results, one can then conclude on optimal values for $\beta'$, $t'$, and $t''$ in the evaluation of the response function according to Eq.~\eqref{eq:chi_schemeFull}. For the demonstrational purposes of this paper, it is sufficient to not explore the full three-dimensional parameter space, but to constrain ourselves in the following to the subspace with $t'=t''$, i.e., to the schemes
\begin{align}\nonumber
	\chi^{\beta't'}_{\hA\hB}(\beta,t)
	&= \frac{1}{Z_\beta}\Tr\left(\big[e^{-\beta'\hH}\hB(t')\big] \big[\hA(t'-t) e^{-(\beta-\beta')\hH}\big]\right)\\
	\label{eq:chi_schemeFull2}
	&= \frac{1}{Z_\beta}\Tr\left(\big[e^{i\hH t'}e^{-\beta'\hH}\hB e^{-i\hH t'}\big] \big[e^{-i\hH (t-t')}\hA e^{-(\beta-\beta')\hH}e^{i\hH(t-t')}\big]\right).
\end{align}

\begin{figure}[p]
\includegraphics[width=1\textwidth]{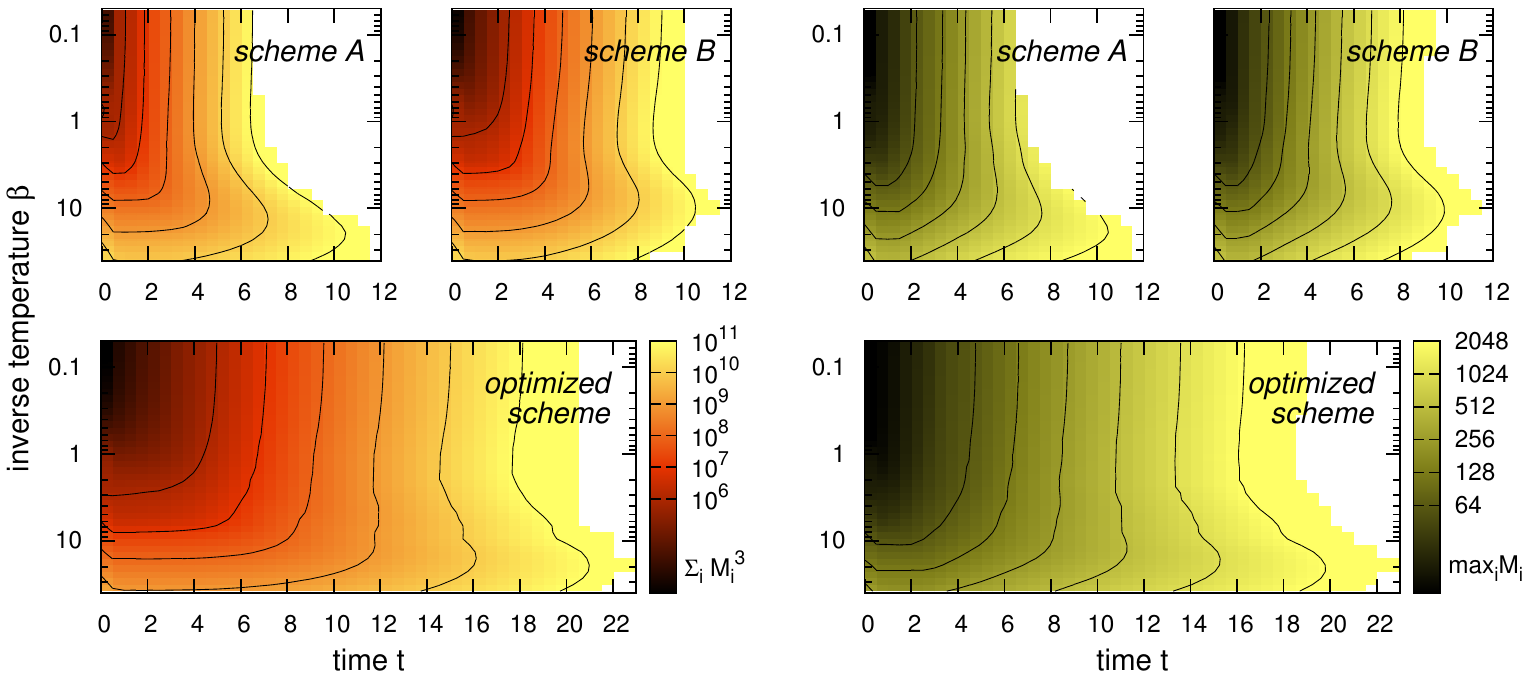}
\caption{\label{fig:cmp_Sp63}As this direct comparison for $J_z=1$ and $\hA=\hB^\dag=\hS^+_{L/2}$ shows (all time axes have the same scales), the maximum reachable times in the optimized scheme exceed those of the earlier schemes roughly by a factor of two. The efficiency measure for the optimization was chosen to be the computation cost per time step (left) and the maximum bond dimension (right), respectively. The truncation weights were chosen as $\epsilon_\beta=10^{-12}$, $\epsilon_t=10^{-10}$.}
\end{figure}

\begin{figure}[p]
\includegraphics[width=1\textwidth]{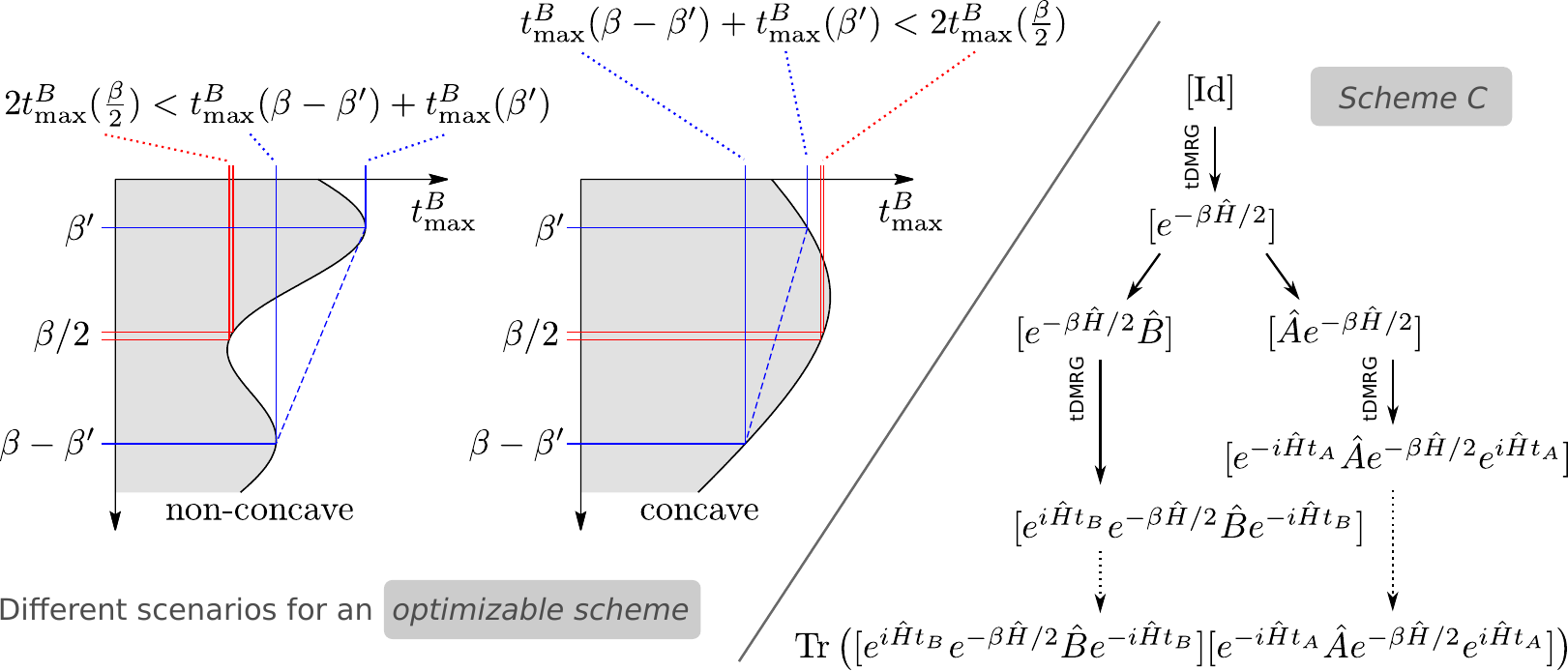}
\caption{\label{fig:schemeOpt} Left: Two possible scenarios for the maximum times $t^\opt_\max$ [Eq.~\eqref {eq:tMaxOpt}] reachable with the optimizable tDMRG scheme \eqref{eq:chi_schemeFull2} for the case $\hA=\hB^\dag$ and given computational resources. If $t^B_\max(\beta')$ is non-concave on the interval $[0,\beta]$, the optimal scheme corresponds to some nontrivial $\beta'$. This can for example occur for systems with dynamics on different energy scales. If $t^B_\max(\beta')$ is concave, the optimal scheme is given by $\beta'=\beta/2$, i.e., \emph{scheme C}. In the examples studied here, $t^B_\max(\beta')$ was found to be almost concave in the relevant temperature ranges. Right:
\emph{Scheme C} for the evaluation of the response function according to Eq. \eqref{eq:chi_schemeC}.}
\end{figure}

Figure~\ref{fig:opt_Sp63Spk0} shows, with $\hB^\dag=\hA$ \cite{foot:A_equal_Bh}, the computation cost per time step and the maximum bond dimensions for the optimized scheme. The maximum reachable times $t^\opt_\max$ of the optimized scheme are at least twice as large as the reachable times $t^B_\max(\beta)$ for \emph{scheme B}, specifically $t^\opt_\max(\beta)\gtrsim 2 t^B_\max(\beta/2)$. See also Figure~\ref{fig:cmp_Sp63}. The factor $1/2$ in the temperature argument is inessential as one finds that $t_\max(\beta)\approx t_\max(\beta/2)$. This is due to the fact that $t_\max$ varies slowly as a function of $\log\beta$ for all response functions considered in this paper. Let us discuss the possible scenarios for the optimum values of $\beta'$ and $t'$ for the case $\hB^\dag=\hA$: The computation cost for \emph{scheme B} is dominated by the tDMRG computation of the operator $\big[e^{-i\hH t}\hA e^{-\beta'\hH/2}e^{i\hH t}\big]$ in Eq.~\eqref{eq:chi_schemeB}. The corresponding maximum time, reachable for some given computational resources, truncation weight $\epsilon$, and inverse temperature $\beta'$, is denoted by $t^B_\max(\beta')$. The maximum reachable time for the scheme \eqref{eq:chi_schemeFull2} as a function of $\beta$ is then given by
\begin{equation}\label{eq:tMaxOpt}
	t^\opt_\max(\beta) = \max_{0\leq \beta'\leq\beta}\left( t^B_\max(\beta') + t^B_\max(\beta-\beta') \right).
\end{equation}
There are two possible scenarios, as depicted in Figure~\ref{fig:schemeOpt}. If $t^B_\max(\beta')$ is concave on the interval $[0,\beta]$, the optimal scheme corresponds to $\beta'=\beta/2$. Otherwise, there will be an optimum $\beta'\neq\beta/2$. In the examples studied here, $t^B_\max(\beta')$ was found to be almost concave in the relevant temperature ranges.

One can hence devise a corresponding \emph{scheme C} that is near-optimal among the schemes \eqref{eq:chi_schemeFull2} (at least for $\hB^\dag=\hA$), does not require any optimization, but outperforms \emph{scheme B} by a factor of two in the maximum reachable times. It is only outdone by \emph{scheme A} at very low temperatures. The scheme corresponds to the choice $\beta'=\beta/2$ in Eq.~\eqref{eq:chi_schemeFull2}
\begin{align}\nonumber
	\chi^C_{\hA\hB}(\beta,t_A+t_B)
	&= \frac{1}{Z_\beta}\Tr\left(\big[e^{-\beta\hH/2}\hB(t_B)\big] \big[\hA(-t_A) e^{-\beta)\hH/2}\big]\right)\\
	\label{eq:chi_schemeC}
	&= \frac{1}{Z_\beta}\Tr\left(\big[e^{i\hH t_B}e^{-\beta\hH/2}\hB e^{-i\hH t_B}\big] \big[e^{-i\hH t_A}\hA e^{-\beta\hH/2}e^{i\hH t_A}\big]\right).
\end{align}
After the imaginary-time evolution that yields $[e^{-\frac{\beta}{2}\hH}]$, one runs two real-time tDMRG simulations to obtain MPOs $\big[e^{i\hH t_B}e^{-\frac{\beta}{2}\hH}\hB e^{-i\hH t_B}\big]$ and $\big[e^{-i\hH t_A}\hA e^{-\frac{\beta}{2}\hH}e^{i\hH t_A}\big]$. With Eq.~\eqref{eq:chi_schemeC} one then obtains $\chi^C_{\hA\hB}(\beta,t_A+t_B)$; see Figure~\ref{fig:schemeOpt}. The accuracy of the MPOs should be kept under control during the whole simulation, for example, as described in Section~\ref{sec:DMRG} by bounding the truncation error \cite{foot:commute}. If this is done properly, it is of minor importance what specific $t_A$ and $t_B=t-t_A$ are chosen to evaluate $\chi_{\hA\hB}(\beta,t)$ for a given time $t$. The maximum reachable time $t$ is of course given by the sum of the maximum reachable $t_A$ and $t_B$. For the case $\hA=\hB^\dag$ (or cases where $\hB^\dag$ is for example simply a translate of $\hA$ \cite{foot:A_equal_Bh}), the maximum reachable times for $t_A$ and $t_B$ are equal, and the total maximum reachable time with this scheme is then twice as large as the maximum time of \emph{scheme B}. This makes many more physical applications accessible.

As pointed out above, \emph{scheme C} is optimal among the classes of schemes corresponding to Eq.~\eqref{eq:chi_schemeFull2} if $t^B_\max(\beta')$ is a convex function. For the cases studied here it was indeed found to be convex or almost convex in the considered temperature ranges. One can expect a different behavior for example for systems with dynamics on different energy scales. In such cases one can achieve considerable performance improvements by optimizing among the class of schemes \eqref{eq:chi_schemeFull} or \eqref{eq:chi_schemeFull2}. In exceptional cases it also occurs that evolved operators $\hA(t)$ have an MPO representation with a time-independent bond dimension. This implies $t^B_\max(0)=\infty$ and one can compute the response function for arbitrary times by using $\beta'=\beta$ and $t'=0$ in Eq.~\eqref{eq:chi_schemeFull2}. One such case is the operator $\hS_j^z(t)$ for the XY model ($J^z=0$) which can for all times be written as an MPO of bond dimension $M_j=4$ \cite{Hartmann2009-102}.

\section{\texorpdfstring{\lowercase{t}}{t}DMRG versus TMRG contractions}\label{sec:tDMRG_vs_TMRG}
The time evolution of matrix product states or density operators, as employed in the presented evaluation schemes for thermal response functions, can be implemented within the tDMRG framework in several different ways. The results for the computation costs of the different evaluation schemes are essentially independent of the chosen time evolution algorithm. The currently most common choice \cite{Vidal2003-10,White2004,Daley2004} employs a Trotter-Suzuki decomposition \cite{Trotter1959,Suzuki1976,Suzuki1985-26,Yoshida1990,Hatano2005}. Alternatively, one can use for example the Arnoldi method, Runge-Kutta method or other Krylov subspace approaches \cite{Feiguin2005,Garcia-Ripoll2006-8,Wall2012-14}.

\begin{figure}
\includegraphics[width=0.8\textwidth]{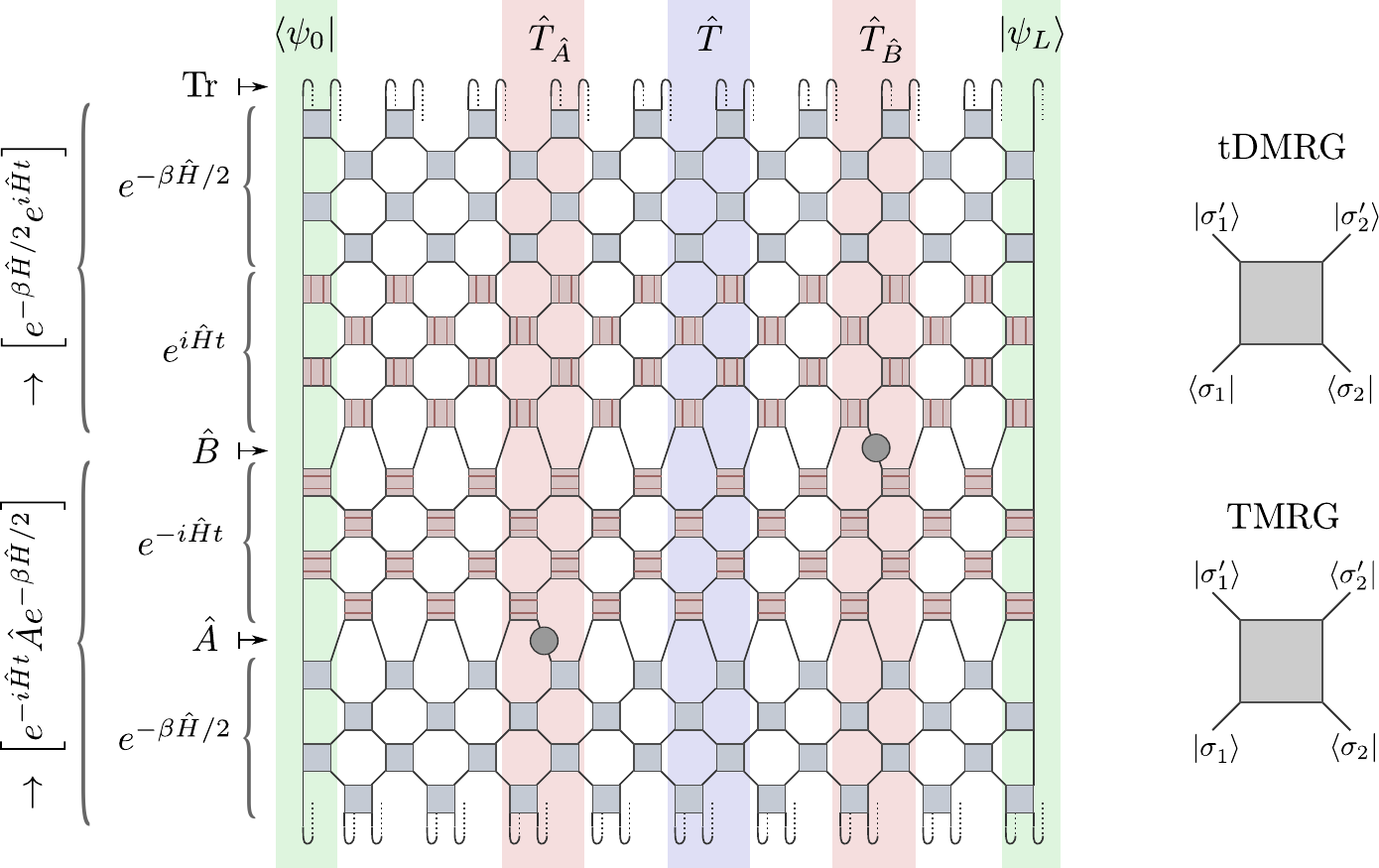}
\caption{\label{fig:tDMRG_vs_DMRG}Visualization of the 2D tensor network that has to be contracted for the evaluation of the thermal response function $Z^{-1}_\beta\Tr(e^{-\beta \hH}e^{i\hH t}\hB e^{-i\hH t}\hA)$ on the basis of \emph{scheme A} when using a Trotter-Suzuki decomposition \eqref{eq:Trotter} of the operator exponentials (nearest-neighbor interactions). This can be done either using tDMRG or using TMRG. In the tDMRG method, one contracts one row of local gates after another (with intermediate truncation steps), evolving two MPOs that are oriented horizontally. The final evaluation of the response function according to Eq.~\eqref{eq:chi_schemeA} corresponds to taking the Hilbert-Schmidt scalar product of the two MPOs. Alternatively, one can contract this tensor network with TMRG. In this case the meaning of states and dual states is changed as indicated on the right side of the figure. One starts on the left and on the right with vertically oriented MPS. One column of local gates after another, the transfer matrices, is applied to those MPS, again with intermediate truncation steps to reduce the bond dimensions. Finally, the expectation value is obtained by evaluating the scalar product of the two MPS.}
\end{figure}

Let us consider in the following a Hamiltonian $\hH=\sum_{i=1}^L\hat{h}_i$ with nearest-neighbor interactions $\hat{h}_i$, operators $\hA$ and $\hB$ with finite spatial support (for simplicity single sites), and implementing the time evolution with a Trotter-Suzuki decomposition. This case allows for an alternative evaluation of the thermal response functions on the basis of the transfer matrix renormalization group (TMRG) \cite{Nishino1995-64,Bursill1996-8,Shibata1997-66,Wang1997-56}.
The Trotter-Suzuki decompositions of real- or imaginary-time propagators read
\begin{align}
	e^{-\tau\hH}&=\left(e^{-\Delta\tau(\hH_\odd+\hH_\even)}\right)^N\nonumber\\
	&=\left(e^{-a_{m+1}\Delta\tau\hH_\odd}\prod_{n=1}^m\left(e^{-b_{n}\Delta\tau\hH_\even}e^{-a_{n}\Delta\tau\hH_\odd}\right) + \mc O\left((\Delta\tau)^{p+1}\right) \right)^N
	\label{eq:Trotter}
\end{align}
where $\tau=\beta$ or $\tau=\pm it$, respectively, $\Delta\tau=\tau/N$ is the time step, and $\hH_{\odd,\even}$ contain the Hamiltonian terms on even and odd bonds, respectively. The coefficients $a_n$ and $b_n$ sum up to one ($\sum_n a_n,\sum_n b_n=1$) and define different decompositions of order $p$ and number of stages $m$ per time step. A common choice is the leapfrog algorithm which has $m=1$ stages and order $p=2$ with $a_1=a_2=1/2$ and $b_1=1$. In this work, a decomposition with $m=5$ stages and order $p=4$ was used.
When the operator exponentials occurring in the formula
$Z^{-1}_\beta\Tr(e^{-\beta\hH/2} e^{i\hH t}\hB e^{-i\hH t}\hA e^{-\beta\hH/2})$
for the response function in \emph{scheme A}, Eq.~\eqref{eq:chi_schemeA}, are decomposed by such a Trotter-Suzuki decomposition, one ends up with the task of contracting a 2D tensor network (quantum cellular automaton) of local gates as displayed in Figure~\ref{fig:tDMRG_vs_DMRG} with the space direction being horizontal and the time direction being vertical. Similar tensor networks result for the other evaluation schemes.

The tDMRG approach \emph{scheme A} corresponds to starting with a trivial MPO $[\id]$ that represents the identity with bond dimensions $M_i=1$ $\forall_{i\in [1,L]}$. Beginning at the bottom, one then contracts one row of local gates after another to that MPO until reaching the row that contains operator $\hB$, yielding $[e^{-i\hH t}\hA e^{-\beta \hH/2}]$. Multiplying, in every step of this iteration, one row of gates with the current MPO in an exact manner, gives a resulting MPO with increased bond dimensions. A subsequent truncation step reduces the bond dimensions again. The degree to which the bond dimensions are reduced in the truncation steps is controlled by the predefined truncation weight \eqref{eq:MPOtrunc} which determines the accuracy of the computation. Similarly, starting from the top, one contracts rows of gates to obtain $[e^{-\beta \hH/2}e^{i\hH t}]$. Finally, one computes the response function by evaluating the Hilbert-Schmidt scalar product \eqref{eq:chi_schemeA} as visualized in Figure~\ref{fig:mpo}.

In the TMRG approach, the columns of local gates in Figure~\ref{fig:tDMRG_vs_DMRG} are interpreted as transfer matrices $\hat T=\hat T(\beta,t)$ (also $\hat T_{\hA}$ and $\hat T_{\hB}$) and operate on a Hilbert space of dimension $d^{\beta/\Delta\beta + 2t/\Delta t}$, where $d$ denotes the dimension of the single-site Hilbert space. The last column corresponds to a pure state $|\psi_L\ket$ and the first column to a dual state $\bra\psi_0|$. Those transfer matrices and states have matrix product representations with bond dimensions $M_i\leq d^2$ $\forall_i$. Starting from the left and the right with the states $\psi_{0}$ or $\psi_{L}$, respectively, one transfer matrix after another is applied with intermediate truncation steps, for example, until one reaches the column containing operator $\hB$. The response function is then given by the overlap of the two resulting MPS,
	$[\bra\psi_0|\hat T\dotsb\hat T\hat T_{\hA}\hat T\dotsb\hat T][\hat T_{\hB}\hat T\dotsb\hat T|\psi_L\ket]$.

The advantages of the tDMRG approaches, pursued in this work, are that, (i), the number of matrices in the MPOs is fixed by the lattice size $L$ instead of growing linearly with the time $t$ and inverse temperature $\beta$ in the TMRG approach, (ii), the operators $\hA$ and $\hB$ can have non-local support, (iii), one can evaluate the spectral function for all times $t$, up to the maximum reachable time, by two tDMRG runs whereas, in the TMRG approach, one has to do a new calculation for every time $t$ of interest.

Concerning (ii), it should be pointed out that non-local operators $\hA$ and $\hB$ would also be accessible in the TMRG approach as long as they are MPOs with sufficiently small bond dimensions.
In Refs.~\cite{Sirker2005-71,Sirker2006-73} the focus was on autocorrelation functions of local operators in the thermodynamic limit $L\to\infty$. In this case, one does not need to contract one column after another, but can evaluate the response function directly from the left and right transfer matrix eigenvectors with maximum eigenvalue. When one uses the so-called infinite-system DMRG \cite{White1992-11}  to obtain those eigenvectors (a single build-up sweep for the finite imaginary-time lattice; not to be mistaken with the algorithm presented in Ref.~\cite{McCulloch2008_04} which generates translationally invariant MPS for infinite systems), problem (iii) does not occur and autocorrelation functions can be evaluated for all accessible times $t$ in a single run. One should however be careful in applying infinite-system DMRG only, as this algorithm has several pitfalls \cite{Schollwoeck2005}. At least for larger times, finite-system DMRG (multiple sweeps) should be necessary.

\section{Conclusion}
In this paper, I have studied and explained the computation costs of different tDMRG schemes for the efficient and precise evaluation of finite-temperature response functions of strongly correlated quantum systems. Simplifying the notation to some extent by formulating everything in terms of MPOs, elucidated the effects of quasi-locality on the costs. The new class of optimizable evaluation schemes, Eqs.~\eqref{eq:chi_schemeFull} and \eqref{eq:chi_schemeFull2}, typically outperforms the earlier schemes from the literature in terms of the maximum reachable times by a factor of $\gtrsim 2$. This gives access to many more physical applications. The novel \emph{scheme C}, that requires no additional optimization and is near-optimal in many typical examples, can be expected to be the method of choice for future applications. For more complex models, like systems with dynamics on different energy scales, \emph{scheme C} still outperforms the older tDMRG methods, but one can achieve further substantial performance gains by first studying the costs for the computation of operators $\big[e^{i\hH s}e^{-\alpha\hH}\hB e^{-i\hH s'}\big]$, and then using this information to determine the optimal parameters for the scheme \eqref{eq:chi_schemeFull}.
Finally, it has been argued that the tDMRG schemes are in some respects favorable to corresponding TMRG variants. As a first application, \emph{scheme C} has been used in Ref.~\cite{Barthel2012_12unused} to calculate, for 1D bosons in the quantum critical regime with dynamic critical exponent $z=2$, the universal scaling function for the thermal spectral function.

Discussions with J.\ Barthel, C.\ Karrasch, A.\ Kolezhuk, I.\ P.\ McCulloch, T.\ Nishino, S.\ Sachdev, J.\ Sirker, and U.\ Schollw\"ock as well as financial support through DFG FOR 801 are gratefully acknowledged.

\clearpage
\bibliographystyle{prsty} 

\begin{thebibliography}{10}
\bibitem{foot:commute}
 Please note that the expressions \eqref {eq:chi_schemeA}, \eqref
   {eq:chi_schemeB}, \eqref {eq:chi_schemeFull}, and \eqref {eq:chi_schemeC} for
   the response function in the different evaluation schemes give the same
   values, the exact $\chi _{\hA \hB }(\beta ,t)$ of Eq.~\eqref {eq:chiAB}, only
   if the MPO representations of the operators in square brackets are exact --
   corresponding to exponentially big bond dimensions with $\max_i \log M_i=\mc
   O(L)$. A guiding principle of the DMRG approach is that in typical condensed
   matter applications such exponentially big $M_i$ are not required.
   Corresponding truncations of the MPOs result in slight deviations of the
   values from the different evaluation schemes. It is essential to keep the
   errors \eqref {eq:MPOtrunc} in these MPO truncations controlled and small at
   all times.
\bibitem{foot:A_equal_Bh}
 It might seem restrictive to only provide data for the choice $\hB = \hA ^\dag
   $. However, the cases where $\hB $ is a translate of $\hA ^\dag $ are very
   similar. This covers the typical cases of interest like $\hA =\hS ^+_i$, $\hB
   =\hS ^-_j$ etc. If neither site $i$ nor site $j$ are close to the boundaries
   of the system, the resulting computation costs and maximum bond dimensions
   are very similar to those occurring for $i=j$. The peaks in the bond
   dimensions simply occur at different locations in the lattice, but have
   similar shape and height; compare to Figure~\ref {fig:AB_Sp63_slices}.
\end{thebibliography}

\end{document}